# Neutron irradiation test of depleted CMOS pixel detector prototypes

Igor Mandić[a,*], Vladimir Cindro[a], Andrej Gorišek[a], Bojan Hiti[a], Gregor Kramberger[a], Marko Mikuž[a,b], Marko Zavrtanik[a],
Tomasz Hemperek[c], Michael Daas[c], Fabian Hügging[c], Hans Krüger[c], David-Leon Pohl[c], Norbert Wermes[c],
Laura Gonella[d]

[a] *Jožef Stefan Institute, Jamova 39,Ljubljana, Slovenia*

[b] *University of Ljubljana, Faculty of Mathematics and Physics,
Jadranska 19, Ljubljana, Slovenia*

[c] *Physikalisches Institut, Universität Bonn,
Nußallee 12, 53115 Bonn,Germany*

[d] *School of Physics and Astronomy,University of Birmingham,
Edgbaston, Birmingham, B15 2TT, United Kingdom*

*E-mail:* igor.mandic@ijs.si

ABSTRACT: Charge collection properties of depleted CMOS pixel detector prototypes produced on p-type substrate of 2 k$\Omega$cm initial resistivity (by LFoundry 150 nm process) were studied using Edge-TCT method before and after neutron irradiation. The test structures were produced for investigation of CMOS technology in tracking detectors for experiments at HL-LHC upgrade. Measurements were made with passive detector structures in which current pulses induced on charge collecting electrodes could be directly observed. Thickness of depleted layer was estimated and studied as function of neutron irradiation fluence. An increase of depletion thickness was observed after first two irradiation steps to $1\cdot 10^{13}$ n/cm$^2$ and $5\cdot 10^{13}$ n/cm$^2$ and attributed to initial acceptor removal. At higher fluences the depletion thickness at given voltage decreases with increasing fluence because of radiation induced defects contributing to the effective space charge concentration. The behaviour is consistent with that of high resistivity silicon used for standard particle detectors. The measured thickness of the depleted layer after irradiation with $1\cdot 10^{15}$ n/cm$^2$ is more than 50 μm at 100 V bias. This is sufficient to guarantee satisfactory signal/noise performance on outer layers of pixel trackers in HL-LHC experiments.

---

[*] Corresponding author

# Contents

## 1. Introduction

The possibility to use active pixel detectors (with signal amplification within the pixel cell) [1] produced in a commercial CMOS process [2] in the environment of general purpose experiments after the upgrade of LHC to the HL-LHC foreseen in the next decade [3] received a lot of attention in the particle physics community. Active detectors in commercial CMOS technology offer many advantages over the "classical" hybrid detector used in present LHC experiments [4-7]. Some of the most important are: higher granularity and therefore resolution, less material in the tracking volume reducing scattering and showers, significantly lower cost, availability of multiple large volume vendors etc.

Monolithic CMOS detectors have been successfully used before for detection of charged particles [8-10] but the charge collection in these detectors rely on diffusion which makes them too slow and too sensitive to radiation for HL-LHC. In [11] it was shown that depleted active pixel sensors can be made in a commercial HV CMOS technology which allows application of higher bias voltages resulting in larger depleted depths and therefore significant charge collection from drift of charge carriers which is necessary for HL-LHC environment. This was followed by several other developments in various flavours of CMOS processes resulting in prototypes of active pixel detectors with significant depleted thickness [12-14] which can be commonly named as *depleted CMOS pixels* [15].

Recent irradiation study of depleted CMOS pixels irradiated with reactor neutrons up to HL-LHC fluence was published in [16]. The study was performed on test structures produced on p-type wafers with two initial resistivities. Chip HV2FEI4 [17] was produced in the AMS180 process on 10 Ωcm wafer and CHESS-1 chip [18] was made in AMS350 process and initial resistivity 20 Ωcm. It was found that removal of initial acceptors by neutron irradiation plays an important role up to relatively high neutron fluences. This is different than in the case of high resistivity materials usually used for charged particle detectors but not surprising given the orders of magnitude different concentrations of initial acceptors. Results of another more recent study with samples produced in AMS180 process [19] support the findings from [16].

Very promising and extensively studied depleted CMOS pixel detectors were produced by LFoundry [14] in a 150 nm CMOS process on a p-type substrate of ~2 kΩcm resistivity – a material more typical for ionizing particle detectors. More details about the process, design and performance of LFoundry detector prototypes can be found in [14,20,21]. In this paper we report about measurements of charge collection properties using the Edge-TCT method with LFoundry detector samples irradiated with reactor neutrons up to 1 MeV equivalent fluences of $\Phi_{eq} = 8\cdot10^{15}$ n/cm$^2$.

## 2. Samples and irradiation

Measurements were made with passive detector structures on CCPD_LF chips [20,21]. The schemes of the structures used for this work are shown in Figures 1 and 2. There is no CMOS circuitry implemented in the passive detector therefore it has a structure similar to a standard diode detector. Devices were produced on a ~ 700 µm thick substrate which could be biased through the common p-type ring surrounding all test structures. Selection of samples was thinned to 300 µm or 100 µm and back plane was implanted and metalized allowing electrical connection and therefore biasing the substrate through the back plane. The test structures are surrounded by an n-type ring which can be contacted separately to better define the active volume of the investigated structure. Maximal bias voltage before breakdown was about 100 V and it increased to 150 V after irradiation.

Connecting the highly doped n-type implant to an electrical potential higher than that of the substrate contact forms a depleted region in the substrate. Charge carriers released in the depleted region by passage of charged particle or by a laser pulse move in the electric field in the depleted region inducing the transient current on the electrodes which can be observed on the oscilloscope using a high bandwidth amplifier – this is the so called Transient Current Technique (TCT) [22].

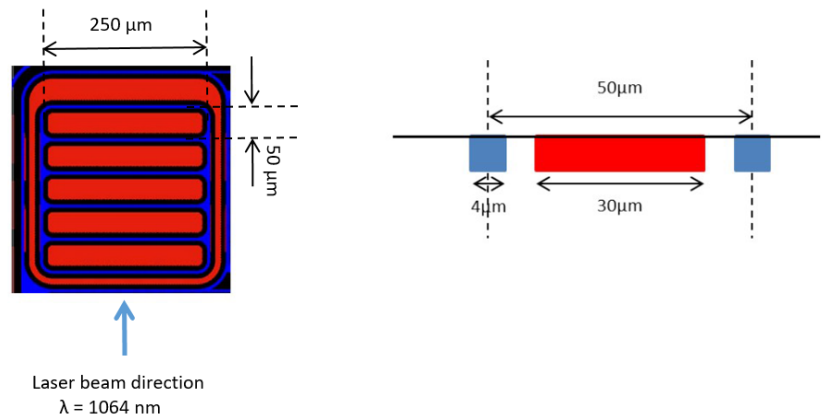

Figure 1. Layout of the passive device comprised of five 50 µm x 250 µm pixels consisting of n-type implants (red) separated by p-type implants (blue, see also the legend in Figure 2). Drawing on the right shows the cross section across the narrow side of the pixel. The five n-type implants are connected together and can be contacted with one wire bond. The n-type ring can be contacted separately. The ring has the asymmetric shape due to the layout of the device on the chip. Direction of laser beam in E-TCT measurements is indicated in the drawing. The p-type substrate containing the implanted structure is not coloured in drawing on the right.

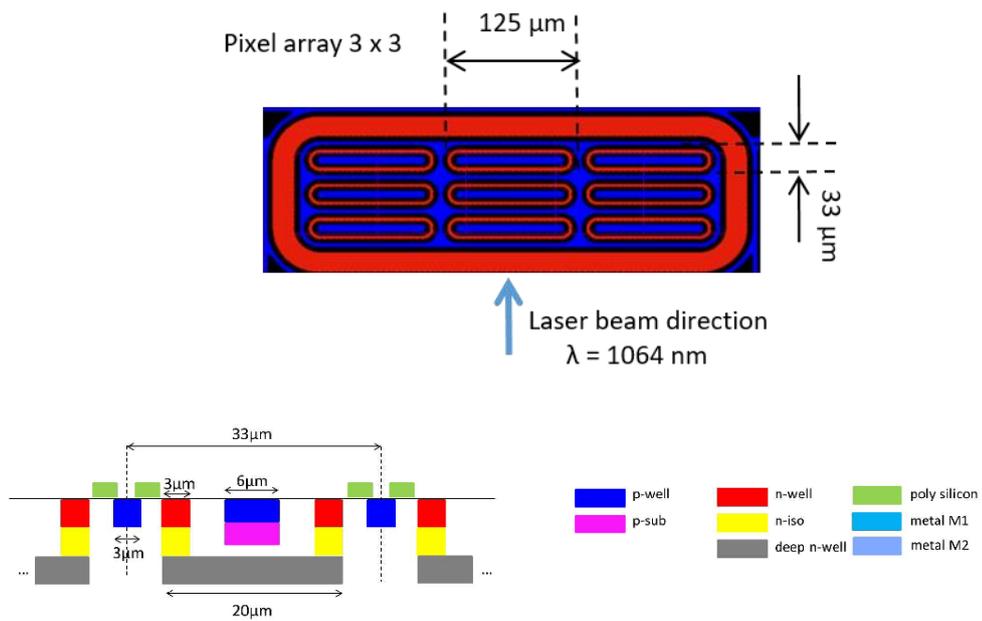

Figure 2. Drawing of 3x3 array of 33 µm x 125 µm pixels. The arrow shows the direction of laser beam in E-TCT. The scheme of the cross section across the narrow side of one pixel is shown below with legend of the colours showing deep n-wels (nw), n-isolation layer (niso), deep n-well (dnw), p-wells (pw), metal (M1), poly silicon resistors (poly-Si) and substrate (psub). The p-type substrate containing the implanted structure (below the line) is not coloured. The pixels (deep n-wells) can be contacted with two separate wire bonds, one for the central pixel and one for the surrounding 8 pixel connected together.

If the side of the device is illuminated with short pulses (< 300 ps, repetition rate 500 Hz) of narrow (FWHM ≤ 10 µm) beam of infra-red ($\lambda$ = 1064 nm) light directed parallel to the surface then this is called the Edge-TCT or E-TCT method [23]. The investigated device is positioned in the beam with sub-micron precision by movable stages so the depth of the substrate at which the charge carriers are released along the laser beam is known [23]. The scheme of the setup and connection scheme of the detector is shown in Figure 3.

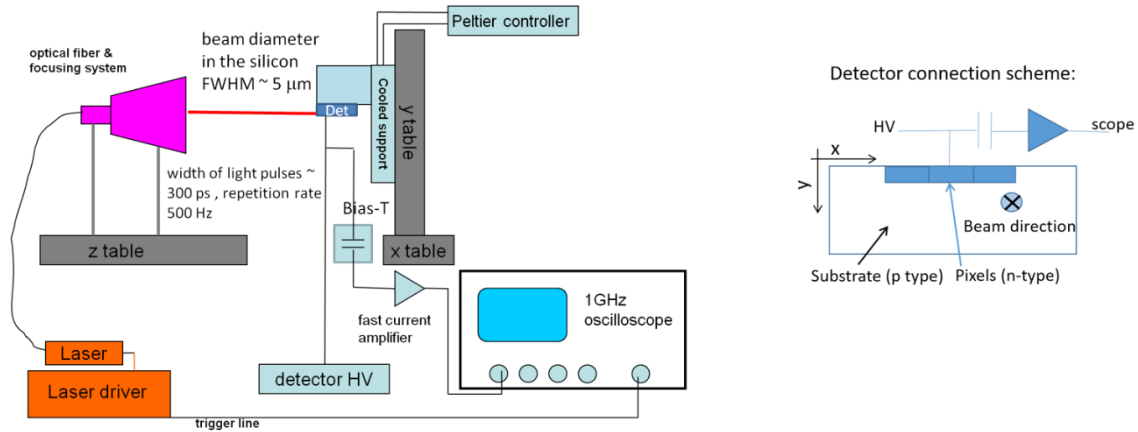

Figure 3. Scheme of the TCT setup and the detector connection scheme. The substrate can be contacted either through the p-type ring implanted on top of the device (not shown in the scheme) or through the back plane for thinned samples with processed back plane.

Irradiations were done in the TRIGA reactor in Ljubljana [24]. Detectors were irradiated with neutrons to fluences ranging from $1\cdot10^{13}$ to $8\cdot10^{15}$ of 1 MeV NIEL equivalent neutrons per cm$^2$. During irradiation to fluences $1\cdot10^{14}$ n/cm$^2$ and larger the reactor power was set to 250 kW which corresponds to neutron flux of $1.7\cdot10^{12}$ $n_{eq}$/cm$^2$/s. For lower fluences the reactor power was set to 25 kW reducing the neutron flux by a factor of ten to increase irradiation time. Fluences were monitored with standardised measurements of leakage current increase in dedicated diodes [25]. Temperature in irradiation tube is about 20°C however it should be noted that during irradiation detectors are heated to over 40 °C by high background radiation [26]. After each irradiation step detectors were kept in the freezer except for few hours needed for transport, mounting and during E-TCT measurements when detectors were at room temperature. Before the first measurement the detectors were at room temperature for max ~ 3 hours. After first measurement detectors were annealed for 80 minutes at 60 °C and measurements were repeated.

Not thinned and 300 μm thick devices were irradiated in steps and after each irradiation step measurement and annealing procedure were repeated while in the case of 100 μm devices different samples were irradiated to different fluences. The study was done with two un-thinned chips, two 300 μm thick devices with processed back plane and twelve 100 μm thick devices, two for each fluence step.

## 3. Measurements of charge collection profiles of the passive device

Measurement and analysis techniques used for this study are similar to E-TCT measurements with CMOS samples described in [16] and the reader may find additional clarifications in that document. The investigated structures are shown in Figures 1 and 2. The structure shown in Figure 1 is an approximation of a pad detector and measurements presented in this paper were made with this type of devices unless otherwise noted. Figure 2 shows a 3x3 pixel array and the main goal of measurement with these structures was to explore their response in the region between neighbouring pixels.

The devices under test were moved in the laser beam in *x* and *y* direction (see Fig 3) in 5 μm steps and at each position 100 pulses were recorded and averaged by a digital oscilloscope. Examples of induced current pulses recorded with laser beam directed at different depths (*y* coordinate) of the unirradiatied detector with *x* positioned at the middle of the pixel are shown in Figure 4a. The pulses are the result of the current induced by the drift of carriers in the weighting field of the detector convoluted with the transfer function of the system and the shape of the laser pulse. It can be seen in Figure 4a that at low depth, close to the surface electrode, the pulse rises steeper and it is shorter than deeper in the detector. In E-TCT the laser pulse releases all carriers at the same depth in the detector so the induced current immediately after the laser pulse is proportional to the sum of carrier velocities at this depth and therefore it reflects the electric field profile in the detector [23]: electric field is higher close to the p-n junction and decreases with the depth in the substrate.

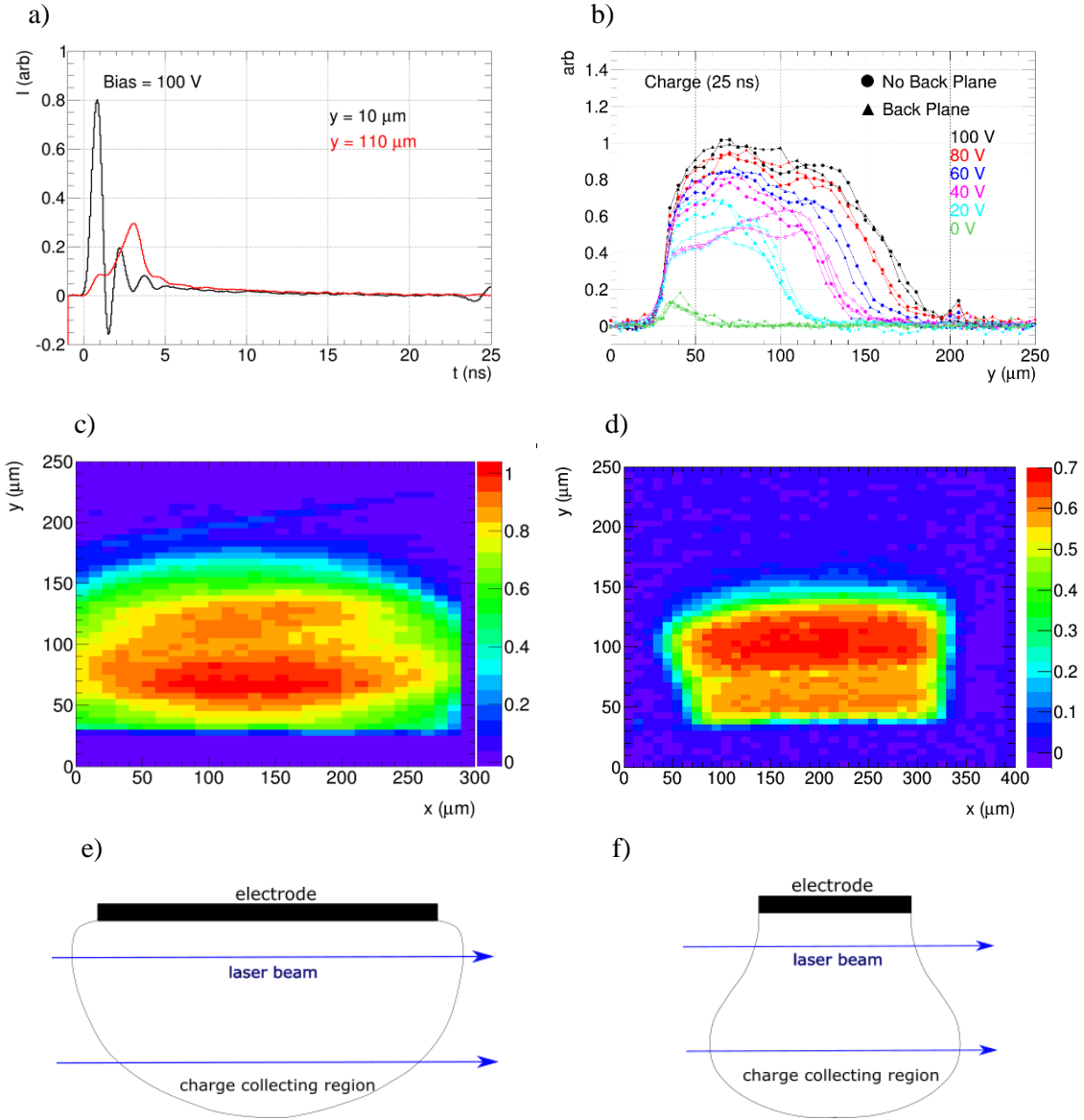

Figure 4: a) current pulses with laser beam at different depth (*y*). The values of coordinate *y* are relative to the surface electrode, b) charge (integral of the induced current pulse in 25 ns) as a function of laser beam depth. The surface of the chip is at $y \sim 40$ μm. The charge vs. depth is shown for devices with (triangular symbols) and without (round symbols) back plane contact at different bias

voltages. Full symbols are measured with n-ring floating, empty symbols and dashed lines are measurements with n-ring biased. The measurements were taken with *x* at the centre of the device. Figure c) shows the charge collection profile in two dimensions at 100 V with n-ring floating, chips surface is at about $y \sim 40$ μm, centre of 250 μm wide pixel is at about $x \sim 130$ μm. Figure d) shows the charge collection profile at 40 V with n-ring connected. Measurements were made before irradiation with structure shown in Fig. 1. The side edges of devices were not polished. Drawing e) sketches the shape of charge collecting region consistent with measurement with n-well ring floating and f) with n-well ring connected.

In this work the integral of the current pulse from 0 to 25 ns is representative of the collected charge. Figure 4b shows the charge collection profiles: the charge as a function of depth across the centre of device. Profiles are shown for different bias voltages for un-thinned device without processed back plane and for the device thinned to 300 μm biased through the back plane. Measurements up to 100 V were made with n-ring (see Figure 1) floating because in this configuration the current was much smaller and higher bias voltages could be applied. With biased n-ring measurements could be done only up to 40 V and are shown in Figure 4b to show how the n-ring affects the profile. When the laser beam is above the surface of the sample (at $y < 30$ μm) the charge is zero. The sharp transition (within 10 μm) from low to high charge is consistent with $\sim 10$ μm wide laser beam entering the device. At bias of 100 V significant charge is measured from $y \sim 40$ μm up to about $y \sim 200$ μm which means about 160 μm deep in the substrate. Initial effective acceptor concentration of p-type silicon with 2 kΩcm resistivity is $6.5 \cdot 10^{12}$ cm$^{-3}$ therefore at 100 V bias the depleted depth in a planar detector should be around 150 μm (see eq. 1) in agreement with the measured depth. Measured curves in figure 4b were aligned offline so that the rising edges of the profiles are at the same *y*. Charge values were scaled to correct for differences in laser beam intensity, edge quality etc... so that the maximal values are about the same at bias of 100 V and the same scaling was used for other bias voltages. There are no significant differences between the charge collection profiles with and without processed backplane. The minor differences in the profile shapes could be a consequence of various sources like surface quality of the edge of the device since it wasn't polished etc.

Figures 4c and 4d show the collected charge in two dimensions with definition of coordinates from Figure 3. Figure 4c shows measurement with floating n-ring and Figure 4d with n-ring connected to the same potential as the readout electrode (but not to the amplifier). It can be seen that n-ring affects the shape of the charge collecting volume below the readout electrode. If n-ring is floating the charge collection region extends laterally also outside of the 250 μm wide n-well and it is "rounded" and so narrower deeper in the substrate. With connected n-ring the charge collecting region matches the electrode size near the electrode and it "stretches" deeper in the oxide. These shapes are similar also in the direction of E-TCT laser beam which reflects in the charge collection profiles. Drawings in figures 4e and 4f qualitatively explain the measured shapes of profiles in Figure 4: collected charge is proportional to the length of laser beam inside the charge collecting region. If n-ring is floating this length is shorter deeper in the substrate while with connected n-ring it initially rises with depth. Due to these effects the measured charge collection profiles deviate from those expected in one dimensional approximation of abrupt junction with uniform space charge concentration in the depleted bulk where the profile would be flat with symmetric transitions on both sides when the laser beam enters (leaves) the area with electric field.

Charge collection profiles measured at 100 V bias with un-thinned and 300 μm thick devices after irradiation are shown in Figure 5a. The profiles in figure 5a are narrower than device thicknesses i.e. the devices are not fully depleted. One can notice a separate narrow charge collection region at $y \sim 330$ μm for the devices with processed back plane. This is due to the electric field formed at the transition between the substrate and the thin highly doped layer at the back plane. It can be seen in Figure 5a that with increasing fluence the charge collection profiles get narrower. This is the consequence of radiation induced increase of effective acceptor concentration and therefore narrower depleted region (eq. 1). Measurements were made up to the fluence of $8 \cdot 10^{15}$ $n_{eq}/cm^2$ and even at this very high fluence the depleted region can clearly be seen. There is no significant difference observed with this experimental method between devices with no back plane processing and the 300 μm samples with processed back plane.

In Figure 5b profiles are shown for 100 μm samples at 70 V bias. At this bias voltage it is expected that devices will be fully depleted. Comparing the charge profile shapes with those in Figure 5a one can clearly see that the transition from high to low charge is sharper, symmetric and it appears at 100 μm from the surface which matches the thickness of the device. As mentioned above, rising edges of profiles were aligned offline which caused somewhat larger spread at the falling edge. The devices remain fully depleted up to the fluence of $1 \cdot 10^{14}$ $n_{eq}/cm^2$ while at the two highest fluence steps the charge profiles start to decrease before the laser beam reaches the back plane. Measurements with two devices at each fluence are shown in Figure 5b except for $\Phi = 10^{13}$ $n/cm^2$ where only one measurements is shown because the laser beam was not well focused during measurement with second device.

As already mentioned, measurements were made immediately after irradiation and repeated after annealing for 80 minutes at 60°C. The profiles shown in Figures 5a and 5b were measured after annealing which increases the width of charge profiles by up to ~ 20% for not fully depleted devices as can be seen in Figures 5c and 5d.

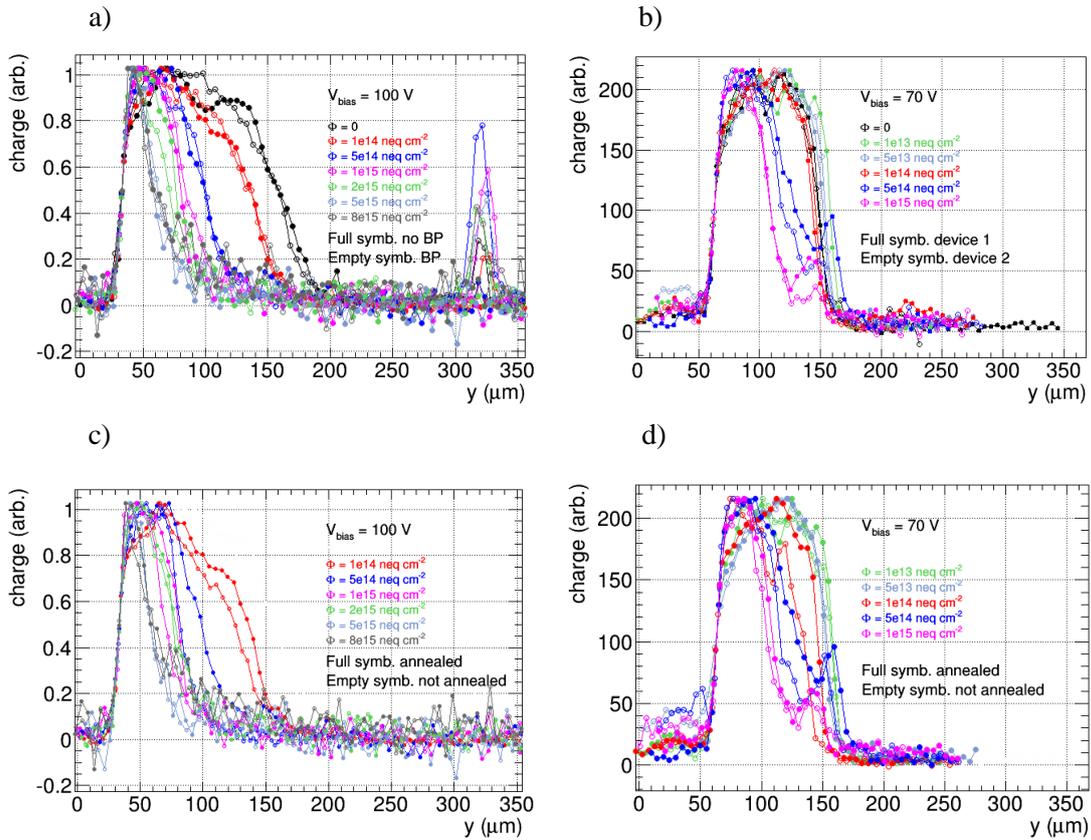

Figure 5. Charge vs. depth after different irradiation fluences indicated in the figure in units of 1 MeV equivalent neutrons/cm$^2$. Figure a) shows un-thinned sample without back plane (full symbols) and sample with processed back plane (empty symbols) thinned to 300 µm. Figure b) shows charge collection profiles of two devices at each fluence thinned to 100 µm with processed back plane. Figure c) shows the comparison of profiles before (empty symbols) and after annealing (full symbols) at different fluences for devices without back plane. Figure d) shows this comparison for 100 µm thick device 1. Measurements were made with the structure type shown in Figure 1.

## 4. Estimation of $N_{eff}$

FWHM of the profile was taken as the measure of the depth of charge collection region. The FWHM is shown as the function of bias voltage for different irradiation fluences for un-thinned and 300 µm thick devices in Figure 6a and for 100 µm devices in Figure 6b.

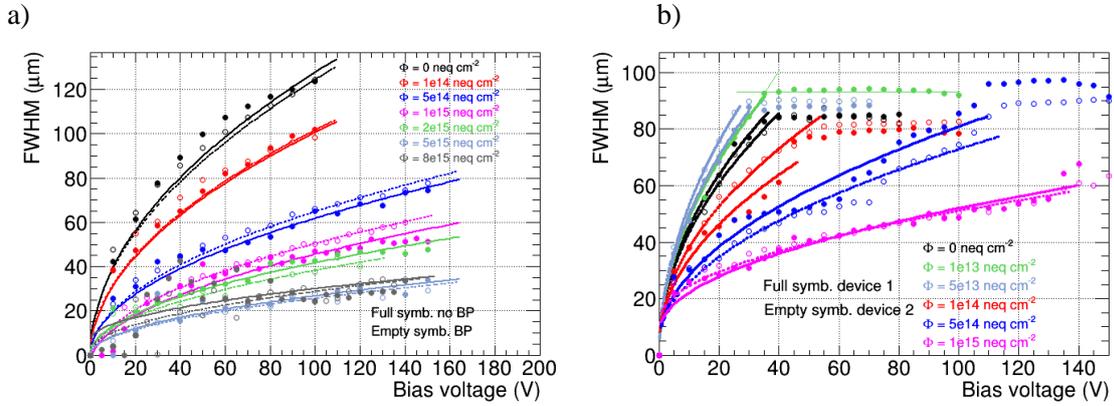

Figure 6. FWHM of the charge collection profile as a function of bias voltage after different irradiation fluences: a) for un-thinned detector without back plane (full symbols) and 300 µm sample with back plane (empty symbols) and figure b) for two 100 µm thick samples at each fluence. Function (2) is fit up to full depletion voltage ($V_{fd}$). For the measurement at $\Phi = 10^{13}$ n/cm$^2$ the thin lines show how $V_{fd}$ was estimated from their cross section.

Before full depletion the scaling of depleted depth $w$ with bias voltage can be described with equation (1):

$$w(V_{bias}) = \sqrt{\frac{2\varepsilon\varepsilon_0}{e_0 N_{eff}} V_{bias}} \qquad (1)$$

where $N_{eff}$ is the effective space charge concentration, $V_{bias}$ the bias voltage, $e_0$ the elementary charge, $\varepsilon_0$ the dielectric constant and $\varepsilon$ relative permittivity of silicon. This relation is valid in planar geometry in approximation of abrupt junction. However, the dependence of depth of charge collection region on bias voltage measured with E-TCT shown in Figure 6 could not be fit well with equation (1). Depleted depth given by equation (1) vanishes at zero bias voltage while in measurements there is an offset and the measured points could be better fit if equation (1) was modified by adding a constant:

$$w = w_0 + w(V_{bias}) \qquad (2)$$

where $w_0$ is the width of the charge collection profile at $V_{bias} = 0$ V and $w(V_{bias})$ is given by equation (1). Parameter $w_0$ can be justified because an offset at low voltage is expected due to a finite laser beam width [16,19] and because of the charge collected by diffusion. Reflections of laser beam from the surface can also affect measurement at shallow depths. Also, the abrupt junction approximation may not be adequate at low bias voltages and linear or more complex doping profiles would be needed to correctly describe the behaviour. But within the precision of experiment described here and the scope of this work these effects can be put aside by adding the ad-hoc parameter $w_0$. The model in equation (2) with $w_0$ and $N_{eff}$ being free parameters fits the data well with some deviations at lower bias for higher fluences as can be seen in Figure 6a. In Figure 6b the measured charge collection width increases up to a knee at a certain voltage after which it is flat for all measurements except for the highest fluence. This knee occurs when the device is fully depleted and $N_{eff}$ can be estimated from the full depletion voltage $V_{fd}$ using relation (1) if $w$ is substituted by the device thickness and $V_{bias} = V_{fd}$. The value of $V_{fd}$ was estimated from Figure 6b as the intersection of horizontal and inclined lines. Horizontal line was a constant fit to the flat part of the measured curve. The inclined line was defined by the first two measured points going towards lower voltage which were more than 5% away from the horizontal line. Values of $N_{eff}$ returned by fitting (2) up to $V_{fd}$ and those calculated from $V_{fd}$ are compatible. Up to $V_{fd}$ measured charge profile widths should not depend on the device thickness. Comparing measurements on figure 6a and 6b this is true for fluences $10^{14}$ n/cm$^2$ and $10^{15}$ n/cm$^2$ while for $5 \cdot 10^{14}$ n/cm$^2$ there is a more significant difference. No specific cause for this difference was recognized so it was treated as device-to-device variation.

## 5. Fluence dependence of $N_{eff}$

Effective space charge concentration $N_{eff}$ as a function of neutron fluence is shown in Figure 7 together with results of a similar study with detectors with much lower initial resistivity produced by AMS [16]. Points in the graph are average values measured with all devices used for this study. For each sample and fluence $N_{eff}$ was estimated by the fit of equation (2) as shown in Figure 6. $N_{eff}$ before irradiation roughly corresponds to initial resistivity of 2 kΩcm. One can see in Figure 7 that for the two lowest fluence points space charge concentration is lower than the initial which is explained by initial acceptor removal. Interaction of initial shallow acceptors (boron) with defects in silicon crystal lattice caused by hadron irradiation can change the properties of the boron substitutional atom [27] so that it doesn't contribute to the space charge. At the same time neutron irradiation also introduces new deep acceptors into silicon which contribute to space charge. However, if the number of removed shallow acceptors is larger than the number of introduced deep acceptors the effective space charge concentration will decrease as a function of fluence in a certain range.
The evolution of $N_{eff}$ with fluence can be described by equation (2) [16,25]:

$$N_{eff} = N_{eff0} - N_c \cdot (1 - \exp(-c \cdot \Phi_{eq})) + g \cdot \Phi_{eq} \qquad (2)$$

where $N_{eff0}$ is the initial acceptor concentration of the substrate, $N_c$ the concentration of the removed acceptors, $c$ the removal constant and $g_c$ the introduction rate of stable deep acceptors for irradiation with neutrons [28]. It was assumed that the short term annealing of $N_{eff}$ was completed, while the long term annealing only takes place on much longer time scales. The free parameters were extracted from the fit of (2) to measurements shown in Figure 7: $c = 10 \cdot 10^{-14}$ cm$^2$, ratio $N_c/N_{eff0} = 0.6$ and $g_c = 0.047$ cm$^{-1}$ with uncertainties of around 30%.

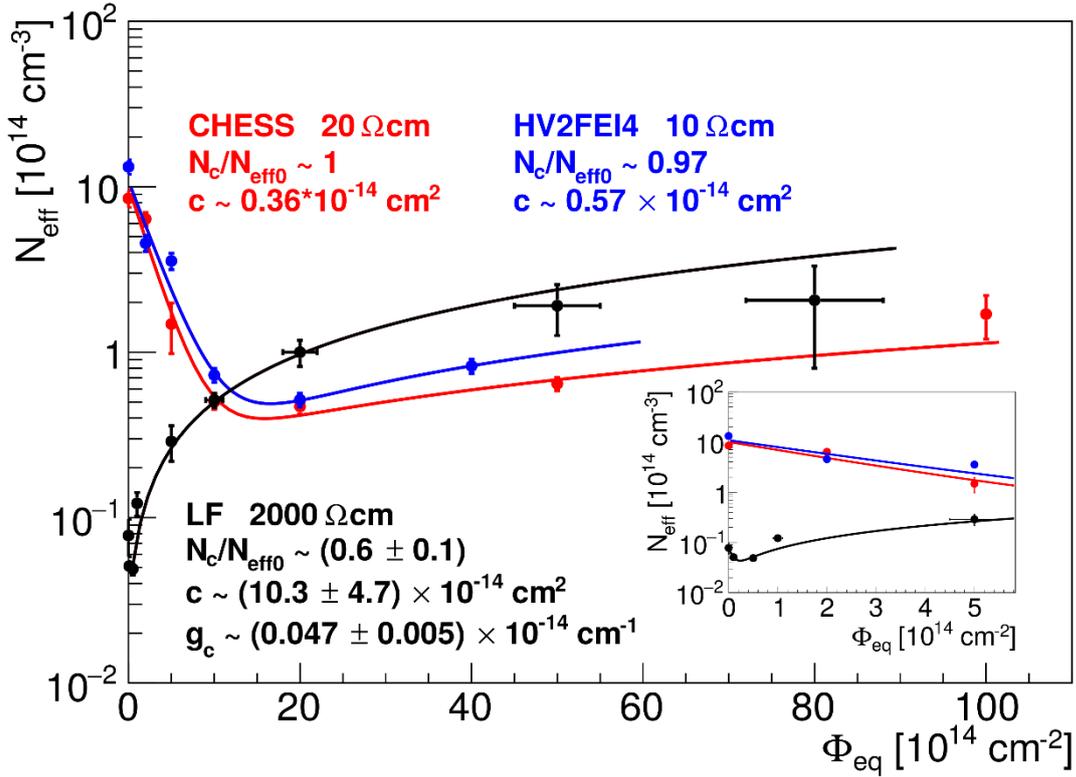

Figure 7. Effective space charge concentration dependence on neutron fluence. The curves are the results of the fit of equation (3) to the measured points. LF are measurements presented in this work and HV2FEI4 and CHESS are 10 and 20 Ωcm initial resistivity samples produced by AMS from [16]. Zoomed view to low fluences is shown in the insert.

Removal constant $c$ is over one order of magnitude larger than the constant measured with a similar method on depleted CMOS pixel detector samples with 10 and 20 Ωcm initial resistivity produced by AMS [16]. The value is close to $c = 1.98 \cdot 10^{-13}$ cm$^2$ reported in [27] from measurements with neutron irradiated detectors made on p-type FZ silicon wafers of similar initial resistivity and as such in agreement with observations from [16,29] that the acceptor removal constant is smaller for material with higher initial acceptor concentration. Recently reported results of similar measurements on CMOS samples with initial resistivities of few hundred Ωcm also fit into this picture [30,31]. They show the growth of depleted depth with irradiation up to fluences larger than in LF (2 kΩcm) but smaller than in CHESS-1 or HV2FEI4 (10 and 20 Ωcm) which would translate in the acceptor removal constant value between those of LF and of CHESS-1 and HV2FEI4. It is important to mention that at very high fluences when acceptor removal is finished $N_{eff}$ is similar in all materials regardless of initial resistivity. The choice of initial resistivity should therefore be governed by the expected performance of readout electronics at lower fluences.

The mechanism behind the dependence of acceptor removal constant on initial resistivity is yet to be understood.

The value of $g_c = 0.047$ cm$^{-1}$ observed in these measurements is by more than a factor of two larger than the value $g_c \sim 0.02$ cm$^{-1}$ measured in MCZ and FZ p-type detector material with similar or higher initial resistivities [28] irradiated with neutrons. But it can be seen that at very high fluence end measured points are departing from the fitted function and the values of $N_{eff}$ are more consistent with $g_c \sim 0.02$ cm$^{-1}$, which is the expected value of this parameter.

## 6. Measurements with pixel array

In addition to depleted depth, the lateral dimension of the region with electric field is also very important for application as segmented detector. E-TCT measurements with pixel arrays are a good tool to explore the uniformity of charge collection near edges of the pixels. In Figure 8 a two dimensional E-TCT scan of a pixel array (see Figure 2) before irradiation and after irradiation to highest fluence can be seen. Measurements were made with all 9 pixels (pixel size: 125 μm x 33 μm) of the array connected to the readout amplifier. The borders of the three pixel rows can be recognized and are more pronounced after irradiation but there are no areas with very low charge collection efficiency. One can note that before irradiation the maximum charge is measured deeper in the substrate and not near the surface of the chip. The effect can be qualitative explained by the shape of the charge collecting volume below the readout electrode as discussed in section 3 (see the drawings in Figure 4). Measurements with pixel array were made with n-ring biased. The shape of the charge collecting volume is related also to the ratio between depleted depth and electrode dimension in the direction of the beam in E-TCT. In case of the pixel array the device dimension is ~ 100 μm (3 x 33 μm) and depleted depth before irradiation is also about 100 μm at 40 V. After irradiation with high neutron fluence depleted depth is much smaller and the effect is not seen any more.

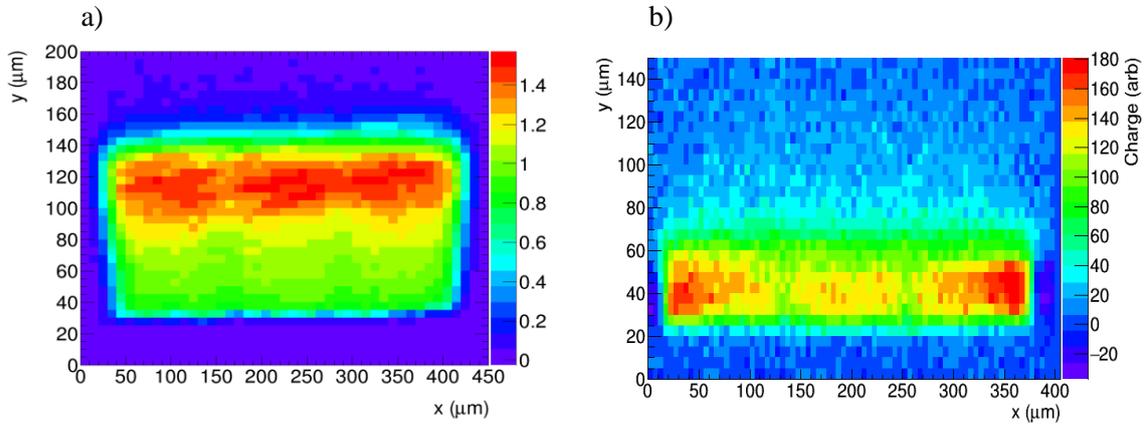

Figure 8. Collected charge measured with E-TCT on the 3x3 pixel array. Coordinate $y$ is the depth of the detector and the chip surface is at $y \sim 20$ μm. Figure a) shows measurement before irradiation at 40 V bias and b) after $8 \cdot 10^{15}$ n/cm$^2$ at 100 V.

## 7. Conclusions

In this work we presented an irradiation study on a set of passive pixel detectors (diodes) produced by LFoundry on p-type material with initial resistivity of 2 kΩcm. The samples were irradiated with reactor neutrons up to $8 \cdot 10^{15}$ n/cm$^2$. The thickness of charge collection layer was measured with Edge-TCT. After irradiation to $1 \cdot 10^{13}$ n/cm$^2$ and $5 \cdot 10^{13}$ n/cm$^2$ an increase of charge collection layer thickness was measured due to initial acceptor removal. At higher fluences the thickness at given voltage decreases with increasing fluence because of radiation induced defects which contribute to the effective space charge concentration $N_{eff}$. $N_{eff}$ was estimated from the voltage dependence of charge collection profile width. The acceptor removal parameters were extracted from dependence of $N_{eff}$ on fluence. The behavior measured in this work is consistent with behavior of standard charged particle detectors produced on silicon substrates with similar initial resistivity and supports observations from [16,29,30] that acceptor removal parameters depend on initial resistivity.

Effect of annealing on charge collection profiles was measured and it was found that annealing for 80 minutes at 60°C has a beneficial effect of increasing the charge collection width by up to 20%.

The charge collection profiles were measured also with a pixel array. Good charge collection uniformity was observed without significant charge collection gaps between pixels up to highest fluences.

The measurements reported in this paper confirm that depleted CMOS pixel detectors can sustain high radiation fluences and are therefore suitable for applications in the inner trackers of HL-LHC experiments. The requirement is the operation at signal-to-noise ratio of S/N ~ 20 or higher. At noise levels of N ~200 electrons (a typical value including threshold dispersions) the required collected signal charge from a passage of a MIP should amount to S ~ 4000 electrons. This corresponds to 50 μm depleted depth which can be achieved at 100 V bias voltage up to the fluence of $1 \cdot 10^{15}$ n/cm$^2$ and higher if higher bias can be applied (Fig. 6).

Measurements of dependence of $N_{eff}$ on irradiation fluence (Fig. 7) as well as the results from [29,30] suggest that an optimum initial resistivity of material for depleted CMOS pixel detectors would be in the range between 100 and 1000 Ωcm to benefit from high signal charge at the start of operation and from the increase of depleted depth due to the acceptor removal at fluences below $1 \cdot 10^{15}$ n/cm$^2$ while at at higher fluences the initial resistivity has lost in importance.

**Acknowledgments**

The authors would like to thank the crew at the TRIGA reactor in Ljubljana for help with irradiation of detectors. Part of this work was performed in the framework of the CERN-RD50 collaboration.